\documentclass[aps,preprint,prd,showpacs,nofootinbib]{revtex4}

\usepackage{latexsym}
\usepackage{amsmath}
\usepackage{graphicx}
\usepackage{subfigure}
\usepackage{dcolumn}
\usepackage{bm}
\usepackage{amssymb}
\usepackage{latexsym}
\usepackage{ulem}
\usepackage{color}
\usepackage[colorlinks,linkcolor=magenta,anchorcolor=cyan,citecolor=blue]{hyperref}

\def\be{\begin{equation}}
\def\ee{\end{equation}}
\def\ba{\begin{eqnarray}}
\def\ea{\end{eqnarray}}

\def\nn{\nonumber}
\def\lf{\left}
\def\rt{\right}

\begin{document}

\title{ The Effective Field Theory of nonsingular cosmology: II }

\author{Yong Cai$^{1}$\footnote{caiyong13@mails.ucas.ac.cn}}
\author{Hai-Guang Li$^{1}$\footnote{lihaiguang14@mails.ucas.ac.cn}}
\author{Taotao Qiu$^{2}$\footnote{qiutt@mail.ccnu.edu.cn}}
\author{Yun-Song Piao$^{1,3}$\footnote{yspiao@ucas.ac.cn}}

\affiliation{$^1$ School of Physics, University of Chinese Academy
of Sciences, Beijing 100049, China}
\affiliation{$^2$ Institute of Astrophysics, Central China Normal University, Wuhan 430079, China}
\affiliation{$^3$ Institute of Theoretical Physics, Chinese
Academy of Sciences, P.O. Box 2735, Beijing 100190, China}

\begin{abstract}


    Based on the Effective
    Field Theory (EFT) of cosmological
	perturbations, 
	we explicitly clarify the pathology in
	nonsingular cubic Galileon models and show how to cure it in EFT
	with new insights into this issue.
	With the least set of EFT operators that are capable to avoid instabilities
	in nonsingular cosmologies, we construct a nonsingular model dubbed the Genesis-inflation model, in which a slowly expanding phase (namely, Genesis)
	with increasing energy density is followed by slow-roll inflation.
	The spectrum of the primordial
	perturbation may be simulated numerically, which shows itself a
	large-scale cutoff, as the large-scale anomalies in
	CMB might be a hint for.

\end{abstract}

\maketitle

\section{Introduction}

Inflation is still being eulogized for its simplicity and
also criticized for its past-incompleteness
\cite{Borde:1993xh}\cite{Borde:2001nh}. A complete description of
the early universe requires physics other than only implementing
inflation.

To the best of current knowledge, the inflation scenario will be
past-complete, only if it happens after a nonsingular bounce which is preceded by a contraction
\cite{Piao:2003zm}\cite{Liu:2013kea}\cite{Qiu:2015nha}, or a
slow expansion phase (namely, Genesis phase) with increasing energy density 
\cite{Liu:2014tda}\cite{Pirtskhalava:2014esa}\cite{Kobayashi:2015gga}.
These two possibilities will be called bounce-inflation and Genesis-inflation, respectively.
Besides being past-complete, a bounce-inflation or Genesis-inflation scenario may explain the probable large-scale anomalies in cosmological microwave
background (CMB) \cite{Liu:2013kea}\cite{Liu:2013iha}. The
nonsingular Quintom bounce
\cite{Cai:2007qw}\cite{Cai:2008qw}\cite{Li:2016xjb} (see also \cite{Roshan:2016mbt}), the ekpyrotic
universe \cite{Khoury:2001wf}\cite{Lehners:2008vx}, the Genesis
scenario
\cite{Creminelli:2010ba}\cite{Hinterbichler:2012yn}\cite{Nishi:2015pta}\cite{Nishi:2016ljg},
and the slow expansion scenario
\cite{Piao:2003ty}\cite{Liu:2011ns}\cite{Liu:2012ww}\cite{Cai:2016gjd} have acquired
intensive attention. In classical nonsingular (past-complete)
cosmologies, the Null Energy Condition (NEC) must be violated for
a period.

The ghost-free bounce models
\cite{Qiu:2011cy}\cite{Easson:2011zy}\cite{Qiu:2013eoa}\cite{Cai:2012va}\cite{Wan:2015hya}
have been obtained in cubic Galileon \cite{Nicolis:2008in} and
full Horndeski theory
\cite{Horndeski:1974wa}\cite{Deffayet:2011gz}\cite{Kobayashi:2011nu}. However, recently,
it has been proved in Ref. \cite{Libanov:2016kfc} that there exists a ``no-go" theorem for cubic Galileon, i.e., the gradient
instability ($c_s^2<0$ with $c_s$ being sound speed of the scalar
perturbations) is inevitable in the corresponding models.
See also \cite{Kobayashi:2016xpl} for the
extension to the full Horndeski theory, and \cite{Ijjas:2016vtq}
for the attempts to avoid ``no-go" in Horndeski theory. Relevant
studies can also be found in
\cite{Ijjas:2016tpn}\cite{Kolevatov:2016ppi}\cite{Akama:2017jsa}.


Recently, in Ref. \cite{Cai:2016thi} (see also
\cite{Creminelli:2016zwa}), we dealt with this issue in the
framework of the Effective Field Theory (EFT)
\cite{Cheung:2007st}\cite{Gubitosi:2012hu}\cite{Piazza:2013coa}\cite{Gleyzes:2013ooa},
which has proved to be a powerful tool.
In EFT,  the quadratic action of scalar perturbation could always be written in the form (see \cite{Cai:2016thi} for detailed derivations)
\be
\label{eft_action02}
S^{(2)}_\zeta=\int d^4xa^3c_1\lf[
\dot{\zeta}^2-c_s^2{(\partial \zeta)^2\over a^2} \rt]\,,
\ee where
we have neglected higher-order spatial derivatives of the scalar perturbation $\zeta$, 
the sound speed squared of scalar perturbation \be
c_s^2=\lf({\dot{c}_3\over a}-c_2 \rt)/c_1 \,, \label{C-introduction}\ee 
with the coefficients $c_1$, $c_2$ and $c_3$ being time dependent parameters in general, and $c_1>0$ is needed to avoid the ghost instability. 
The condition for avoiding gradient instability is $c_s^2\sim{\dot{c}_3/ a}-c_2>0$, which is usually
integrated as \be c_3\Big|_{t_f}- c_3\Big|_{t_i}>\int_{t_i}^{t_f}a
c_2 dt\,. \label{integral} \ee The condition of satisfying the
inequality is to have $c_3$ cross 0, which is hardly possible in
models based on the cubic Galileon
[33][34].
However, we found that it can easily be satisfied by applying the EFT
operator $R^{(3)}\delta g^{00}$ (with $R^{(3)}$ being the 3-dimensional Ricci scalar on spacelike hypersurface and $\delta g^{00}=g^{00}+1$), so that the gradient instability can be
cured.


Though the integral approach (\ref{integral}) is simple and
efficient, some details of curing the pathology might actually be
missed. In this paper, based on the EFT, using a ``non-integral
approach", we revisit the nonsingular cosmologies. We begin
straightly with (\ref{C-introduction}), and clarify the origin of
pathology and show how to cure it in EFT with new insights into what is
happening (Sec. \ref{sec2}). To have practice in this clarification, we build
a stable model of the Genesis-inflation scenario by using the
$R^{(3)}\delta g^{00}$ operator (Sec. \ref{gene-inf}). As a
supplementary remark, we discuss a dilemma in the Genesis scenario
(Sec. \ref{dilemma}).


\section{Re-proof of the ``No-Go" and its avoidance in EFT}\label{sec2}


The EFT is briefly introduced in Appendix \ref{EFT}. In the unitary
gauge, the quadratic action of tensor perturbation $\gamma_{ij}$ is
(see \cite{Cai:2016thi} for the derivation of Eqs. (\ref{tensor-action}) to (\ref{gamma}))
\be S^{(2)}_{\gamma}={M_p^2\over8}\int d^4xa^3
Q_T\lf[ \dot{\gamma}_{ij}^2 -c_T^2{(\partial_k\gamma_{ij})^2\over
a^2}\rt]\,, \label{tensor-action} \ee  where $Q_T= f+{2 m_4^2\over
M_p^2}>0$, $c_T^2=f/Q_T>0$, $f$ and $m_4^2$ are coefficients defined in the EFT action (\ref{eft_action}).

The quadratic action of the scalar perturbation $\zeta$ is given by Eq. (\ref{eft_action02}) with
\ba 
&\,&
c_1=\frac{Q_T}{4 \gamma ^2 M_p^2}
 \Big[
   2 M_p^4 Q_T\dot{f} H
   -2 M_p^2 Q_T \left(  2 f M_p^2 \dot{H}+\ddot{f}M_p^2
      -4 M_2^4\right)
   \nn\\&\,&\qquad\qquad\qquad
   -6 \dot{f}M_p^2 m_3^3
   +3 \dot{f}^2 M_p^4
   +3 m_3^6\Big]
    \,, \label{c1}\\
&\,&c_2=f M_p^2\,,\label{c2}
\\&\,&
c_3={aM_p^2\over \gamma} Q_T Q_{\tilde{m}_4} \,, \label{c3}
\\ &\,&\gamma = HQ_T-{m_3^3\over 2M_p^2}+{1\over2}\dot{f}\,, \quad
Q_{\tilde{m}_4}=f+{2\tilde{m}_4^2\over M_p^2}\,, \label{gamma}\ea
where $M_2^4$, $m_3^3$ and $\tilde{m}_4^2$ are coefficients defined in the EFT action (\ref{eft_action}) and they could be time dependent in general.

Only if $c_1>0$ and $c_s^2>0$, the model is free from ghost and gradient instabilities, respectively. In
nonsingular cosmological models based on the cubic Galileon
\cite{Qiu:2013eoa}\cite{Cai:2012va}\cite{Wan:2015hya}, $c_1>0$ is
not hard to obtain, as can be seen from Eq. (\ref{c1}), since the cubic Galileon contributes the
${m_3^3(t)\over2}\delta K\delta g^{00}$ operator in EFT. However,
since $c_3$ is also affected by ${m_3^3(t)\over2}\delta K\delta
g^{00}$ through $\gamma$, $c_s^2<0$ is actually inevitable, as will be demonstrated in the following.

Since $c_1>0$, the requirement of $c_s^2>0$ equals \be
\lf(H\gamma+{\dot Q_T\over Q_T}\gamma+ {\dot Q_{\tilde{m}_4}\over
Q_{\tilde{m}_4}}\gamma-c_T^2{ \gamma^2\over Q_{\tilde{m}_4} }
-{\dot\gamma}\rt) { Q_T Q_{\tilde{m}_4} \over \gamma^2 }>0.
\label{gamma1} \ee Here, $Q_T\neq Q_{\tilde{m}_4}$ is required,
which cannot be embodied by the Horndeski theory
\cite{Horndeski:1974wa}\cite{Deffayet:2011gz}\cite{Kobayashi:2011nu}. Thus whether $c_s^2>0$ or not is
controlled by the parameter set ($H$, $\gamma$, $Q_T$, $c_T^2$,
$Q_{\tilde{m}_4}$).


In the following, with condition (\ref{gamma1}), we will re-prove the ``no-go" theorem for the
cubic Galileon, and clarify how to cure it in EFT.
Different from the proof in \cite{Libanov:2016kfc}\cite{Cai:2016thi}\cite{Creminelli:2016zwa}, the re-proof is directly based on the derivative inequality instead of integrating it, which we called ``non-integral approach".
We assume that
after the beginning of the hot ``big bang" or inflation, $\gamma=H>0$, $\dot
\gamma <0$ and $Q_{\tilde{m}_4}=1$.

\subsection{Case I:  initially $\gamma<0$}

Since initially $\gamma<0$, $\gamma$ has to cross 0 from
$\gamma<0$ to $\gamma>0$ at $t_\gamma$. The analysis below is also
applicable for all cases with $\gamma$ crossing 0 from
$\gamma<0$ to $\gamma>0$.

In the ekpyrotic and bounce models, initially $\gamma=H<0$. In the Genesis
model \cite{Creminelli:2010ba} and slow expansion model
\cite{Liu:2011ns}, $H>0$ during the Genesis, but actually
$\gamma=H-{m^3_3\over 2M_p^2}<0$, as discussed in
Sec. \ref{dilemma}. Both belong to the Case I.

\underline{\it In the cubic Galileon case}, $f=Q_T=Q_{\tilde{m}_4}=1$. Around $t_\gamma$, condition (\ref{gamma1}) is \be
-{\dot\gamma}>0. \ee We see that $c_s^2<0$ is inevitable around
$t_\gamma$, since ${\dot\gamma}>0$. Thus the nonsingular models
based on cubic Galileon is pathological, as first proved by LMR in
\cite{Libanov:2016kfc}.

\underline{\it In the EFT case}, around $t_\gamma$, condition (\ref{gamma1}) requires \be
\lf({\dot Q_T\over Q_T}\gamma+ {\dot Q_{\tilde{m}_4}\over
Q_{\tilde{m}_4}}\gamma -{c_T^2 \gamma^2\over Q_{\tilde{m}_4} }
-{\dot\gamma}\rt)Q_{\tilde{m}_4}>0. \label{gamma5} \ee We might
have $c_s^2>0$, only if (considering only the case where only one of $Q_T$ and $Q_{\tilde{m}_4}$ is modified while the unmodified one is unity)  around $\gamma=0$
\be {{\dot Q}_T\over Q_T}\gamma> {\dot \gamma},
 \label{solution11}
\ee \be {\rm or}\quad  Q_{\tilde{m}_4}<0,  \quad {\rm or}\quad
{{\dot Q}_{\tilde{m}_4}\over Q_{\tilde{m}_4}}\gamma>{\dot
\gamma}+{c_T^2 \gamma^2\over Q_{\tilde{m}_4} } \quad \lf({\rm for
}\quad {Q_{\tilde{m}_4}}\geqslant 0 \rt). \label{solution12} \ee
In solution (\ref{solution11}), at $t_\gamma$, $\gamma=0$ suggests
$Q_T=0$.  Here, since $\gamma=0$ at $t_{\gamma}$, $c_1\sim
{Q_T/\gamma^2}$ diverges.  One possibility of removing this
divergence is that $\gamma\sim (t-t_{\gamma})^p$ and $Q_T \sim
(t-t_{\gamma})^n$ around $t_\gamma$,
with $n\geqslant 2p$ and $p$, $n$ being constants. In Ijjas and Steinhardt's model
\cite{Ijjas:2016vtq}, $\gamma\sim t-t_{\gamma}$ while $Q_T\sim
(t-t_{\gamma})^2 $, which  belongs to this case.

In the bounce model based on the cubic Galileon, Eq. (\ref{gamma})
gives $\gamma=H-{m_3^3\over 2M_p^2}\neq H$. Generally, the NEC is
violated when ${\dot H}>0$, while the period of $c_s^2<0$
corresponds to the phase with $\gamma\simeq 0$ and ${\dot
\gamma}>0$, these two phases do not necessarily coincide, see
Eq. (\ref{gamma}). As pointed out by Ijjas and Steinhardt
\cite{Ijjas:2016vtq}, it is the sign's change of $\gamma$ that
causes the pathology. Here, we  reconfirmed this point.

In solution (\ref{solution12}), if $Q_{\tilde{m}_4}> 0$, at
$t_\gamma$, $\gamma=0$ suggests $Q_{\tilde{m}_4}=0$; while if
$Q_{\tilde{m}_4}< 0$, since $Q_{\tilde{m}_4}=1$ eventually,
$Q_{\tilde{m}_4}$ must cross 0 at $t_{\tilde{m}_4}$ (generally
$t_{\tilde{m}_4}\neq t_{\gamma}$), at which
$\dot{Q}_{\tilde{m}_4}\gamma>c_T^2\gamma^2$ must be satisfied. In
both cases, ${Q}_{\tilde{m}_4}=0$ is required, as proposed by Cai
et.al \cite{Cai:2016thi} and Creminelli et.al
\cite{Creminelli:2016zwa}.

We see again the details of $Q_{\tilde{m}_4}$ crossing 0. In both the
Genesis model and the bounce model, initially ${Q}_{\tilde{m}_4}=1$,
so if $Q_{\tilde{m}_4}< 0$ around $t_\gamma$, ${Q}_{\tilde{m}_4}$
must cross 0 twice. Thus it seems that
$\dot{Q}_{\tilde{m}_4}\gamma>c_T^2\gamma^2$ is hard to
implement. However, with (\ref{C-introduction}) and (\ref{c3}), one always
could solve ${Q}_{\tilde{m}_4}$ for any given $c_s^2$, \be
{Q}_{\tilde{m}_4}={\gamma\over a M_p^2}\int a\lf(c_1c_s^2+c_2\rt)
dt\,, \label{Qm41}\ee  where $Q_T=1$.
\\

\subsection{Case II: $\gamma>0$ throughout}


Since $\gamma>0$ throughout, we must have
${\dot\gamma}\geqslant0$  during some period
initially\footnote{Of course, in Case II, we could also
have $\dot{\gamma}<0$ during some period, but what we focus on is
the period (i.e., $\dot{\gamma} \geqslant0$) where pathologies
appear.}, otherwise $\gamma$ will diverge in the infinite past.

\underline{\it In the cubic Galileon case}, condition (\ref{gamma1}) is \be \quad
H\gamma-\gamma^2-{\dot\gamma}>0\,. \label{gamma2} \ee In the bounce
model, $H<0$ in the contracting phase, and in the Genesis model, $H\sim
0$ in the Genesis phase, both suggest
$H\gamma-\gamma^2-{\dot\gamma}<0$ \footnote{In the case where
$\gamma$ grows from 0 initially, (\ref{gamma2}) is also
obeyed no more.}. Thus $c_s^2<0$ is inevitable in the corresponding
phases, so the nonsingular models based on the cubic Galileon is
pathological.


We see the Genesis model in the cubic Galileon version again in detail. During the slow
expansion (Genesis phase), $H\sim {1/ (-t)^n}$ with the constant $n>1$. Thus
\be \label{eq16} {{\dot \gamma}\over H\gamma}\simeq {{\dot H}\over H^2}\sim
(-t)^{n-1} \gg 1, \ee which implies $H\gamma\ll {\dot \gamma}$.
Thus with (\ref{gamma2}), we see that $c_s^2<0$ is inevitable in the
slow expansion phase. It seems that if $n=1$, $H\gamma\ll {\dot
\gamma}$ might be avoided. 
However, when $n=1$, we have $H=p/(-t)$ and $a\sim {1/
		(-t)^p}$ with constant $p$, thus $a\rightarrow 0$ in the
	infinite past. From (\ref{eq16}), we see that $c_s^2>0$ requires
		$p={H^2/{\dot H}}>1$. Therefore, the universe is singular, or
	from another point of view, it is geodesically incomplete since the
	affine parameter of the graviton geodesics $\int_{t_i}^{t_f} a dt$
	is finite for $p>1$ when $t_i\rightarrow -\infty$.

\underline{\it In the EFT case}, condition (\ref{gamma1}) requires \be \label{gamma6} \lf({\dot
Q_T\over Q_T}\gamma+ {\dot Q_{\tilde{m}_4}\over
Q_{\tilde{m}_4}}\gamma-{c_T^2 \gamma^2\over  Q_{\tilde{m}_4} }
+H{\gamma}-{\dot\gamma}\rt)Q_{\tilde{m}_4}>0\,. \ee
We might have $c_s^2>0$, only if (considering only the case
where either $Q_T$ or $Q_{\tilde{m}_4}$ is
modified) 
\ba
{\dot{Q}_T\over Q_T}>c_T^2\gamma-H+{\dot{\gamma}\over \gamma},
\label{solution21}
\ea
\ba
  {\rm or} \quad {{\dot Q}_{\tilde{m}_4}\over Q_{\tilde{m}_4}}
  <
  {c_T^2\gamma\over Q_{\tilde{m}_4}}-H+{\dot{\gamma}\over \gamma}
 \quad \lf({\rm initially}\,\,\, Q_{\tilde{m}_4}<
0\rt).\label{solution22}
\ea
Generally, $-H{\gamma}+{\dot\gamma}>0$, as in Genesis model and
bounce model. Thus the solution (\ref{solution21}) suggests ${\dot
Q}_T>0$, so that we will have $Q_T=0$ in infinite past. Thus based
on (\ref{solution11}) and (\ref{solution21}), it seems that though
the pathology can be cured by applying $Q_T$, $Q_T=0$ is
inevitable. A model with (\ref{solution21}) has been proposed by
Kobayashi \cite{Kobayashi:2016xpl} ($Q_T\sim {1\over (-t)^{p}}$,
$p>n>1$). During the Genesis $\gamma\sim H\sim {1/ (-t)^n}$,
$n>1$, (\ref{gamma6}) is \be \lf({ {\dot
Q_T/Q_T}}\rt)^{-1}{\dot\gamma\over \gamma}={n/p}<1.
\label{QT21}\ee Initially, $Q_T\sim {1\over (-t)^{p}}= 0$.

In solution (\ref{solution22}), $Q_{\tilde{m}_4}$ must cross
0 at $t_{\tilde{m}_4}$ to $Q_{\tilde{m}_4}>0$, as pointed out by
Cai et.al \cite{Cai:2016thi} and Creminelli et.al
\cite{Creminelli:2016zwa}. Around $t_{\tilde{m}_4}$,
$\dot{Q}_{\tilde{m}_4}> c_T^2\gamma$ must be satisfied.


In (\ref{gamma6}), if $Q_{\tilde{m}_4}>0$ throughout, \ba
  {{\dot Q}_{\tilde{m}_4}\over Q_{\tilde{m}_4}}
  >
  {c_T^2\gamma\over Q_{\tilde{m}_4}}-H+{\dot{\gamma}\over \gamma} \label{solution23} \ea
is obtained. Thus, similar to (\ref{solution21}), we have
$Q_{\tilde{m}_4}=0$ (which definitely requires $\gamma=0$) in the
infinite past. In the Genesis model, $Q_{\tilde{m}_4}\sim 1/(-t)^p$
and $\gamma\sim 1/(-t)^n$ with $p>n$, since
$\dot{Q}_{\tilde{m}_4}/{Q}_{\tilde{m}_4}>\dot{\gamma}/\gamma$.
However, $p>n$ indicates $\dot{Q}_{\tilde{m}_4}<\gamma$ in the
infinite past ($Q_{\tilde{m}_4}=0$), which violates the inequality
(\ref{solution23}). Thus $Q_{\tilde{m}_4}>0$ throughout seems
unworkable.




\begin{table*}[htbp!]
\begin{center}
\begin{tabular}{|c|c|c|}
\hline \multicolumn{3}{|c|} {Nonsingular cubic Galileon models  } \\
\hline
 & Initially {$\gamma<0$} & $\gamma>0$ throughout\\
\hline
 Crossing 0 for $\gamma$?   & $\surd$ & $\times$   \\
\hline $c_s^2<0$ is inevitable (``no-go")? & $\surd $ & $\surd$ \\
\hline Phase with $c_s^2<0$ (Pathological phase) & \quad ${\dot \gamma}>0$ around $\gamma\simeq 0$ \quad & \quad $H\gamma-{\dot\gamma}<\gamma^2$ \quad \\
\hline \multicolumn{3}{|c|} { Curing pathology in EFT} \\
\hline Conditions of $c_s^2>0$  & (\ref{gamma5})  &  (\ref{gamma6}) \\
\hline Applying $Q_T$ & (\ref{solution11}) &  (\ref{solution21}) \\
\hline Applying $Q_{\tilde{m}_4}$ & (\ref{solution12}) &  (\ref{solution22})  \\
\hline \hline

\end{tabular}
 \caption{ Pathology in nonsingular cubic Galileon cosmological models and its cure in EFT by either $Q_T$ or $Q_{\tilde{m}_4}$. }

\end{center}

\end{table*}

%
%
%
%
%

\section{Application to Genesis-inflation}\label{gene-inf}

In this section, we will build a nonsingular model with
the solution (\ref{solution22}), in which the slow-roll inflation is preceded by a Genesis phase. A Genesis phase is a slowly expanding phase originating form the Minkowski vacuum with a drastic violation of NEC, i.e., $\epsilon\ll -1$, thus the energy density is increasing with the expansion of the universe and hence is free from the initial singularity \cite{Creminelli:2010ba}\cite{Hinterbichler:2012yn} (see also \cite{Piao:2003ty}).  As will be shown below,
our model cannot only get rid of the pathology of instability,
but also give rise to a flat spectrum with interesting features at
large scales.

\subsection{The setup of the model}


The action of the model is \ba\label{genesis-action} &\,& S=\int
d^4x\sqrt{-g}\Big[{M_p^2\over 2}R+M_p^2 g_1(\phi)X +g_2(\phi)
X\Box \phi +g_3(\phi)X^2-M_p^4 V(\phi) \nn\\
&\,&\qquad\qquad\qquad\quad  +{\tilde{m}_4^2(t)\over
2}R^{(3)}\delta g^{00} \Big] \,, \ea where
$X=-\nabla_{\mu}\phi\nabla^\mu\phi/2$, $\Box
\phi=\nabla_{\mu}\nabla^{\mu}\phi$, and $\phi$ is a dimensionless
scalar field, so dimensionless are $g_1(\phi)$, $g_2(\phi)$,
$g_3(\phi)$ and $V(\phi)$.

Mapped into the EFT action (\ref{eft_action}), (\ref{genesis-action})
corresponds to \ba &\,& f=1\,,\label{fff}
\\&\,&
\Lambda(t)=
M_p^4 V
-\frac{1}{2} g_2 \dot{\phi }^2 \left(3 H \dot{\phi }+\ddot{\phi} \right)
+\frac{1}{4} g_3 \dot{\phi }^4\,,
\\&\,&
c(t)=
 \frac{M_p^2}{2} g_1 \dot{\phi }^2
-\frac{1}{2} g_2 \dot{\phi }^2 \left(3 H\dot{\phi }-\ddot{\phi} \right)
+\frac{1}{2} g_{2,\phi } \dot{\phi }^4
+\frac{1}{2} g_3 \dot{\phi }^4\,,
\\&\,&
M_2^4(t)=
-\frac{1}{4} g_2 \dot{\phi }^2 \left(3 H \dot{\phi }+\ddot{\phi} \right)
+\frac{1}{4} g_{2,\phi } \dot{\phi }^4
+\frac{1}{2} g_3 \dot{\phi }^4\,,
\\&\,&
m_3^3(t)= -g_2 \dot{\phi }^3 \,,
\\&\,&
m_4^2=0 \,,
\\&\,&
\tilde{m}_4^2\neq0\,.\label{tildem4}
\ea
%


We can get the background equations  \ba \label{eqH} &\,& 3 H^2
M_p^2=\frac{M_p^2}{2} g_1\dot{\phi }^2
  -3 g_2 H \dot{\phi }^3
  +\frac{1}{2}g_{2,\phi } \dot{\phi }^4
  +\frac{3}{4} g_3 \dot{\phi }^4
  +M_p^4 V \,,
  \\&\,&
\dot{H} M_p^2=
  -\frac{M_p^2}{2} g_1 \dot{\phi }^2
  +\frac{3}{2} g_2 H \dot{\phi}^3
  -\frac{1}{2} g_2 \dot{\phi }^2 \ddot{\phi}
  -\frac{1}{2} g_{2,\phi } \dot{\phi }^4
  -\frac{1}{2} g_3 \dot{\phi }^4\,,  \label{dotH}
  \\&\,&
0=
g_1 \ddot{\phi}
+3 g_1 H\dot{\phi }
+\frac{1}{2}g_{1,\phi } \dot{\phi }^2
\nn\\&\,&\qquad
-\frac{9 g_2 H^2 \dot{\phi}^2}{M_p^2}
-\frac{3 g_2 \dot{H} \dot{\phi }^2}{M_p^2}
-\frac{6 g_2 H \dot{\phi } \ddot{\phi} }{M_p^2}
+\frac{2 g_{2,\phi } \dot{\phi }^2 \ddot{\phi}}{M_p^2}
+\frac{g_{2,\phi \phi }\dot{\phi }^4}{2 M_p^2}
\nn\\&\,&\qquad
+\frac{3 g_3 H\dot{\phi }^3}{M_p^2}
+\frac{3 g_3 \dot{\phi }^2 \ddot{\phi} }{M_p^2}
+\frac{3 g_{3,\phi } \dot{\phi }^4}{4M_p^2}
\nn\\&\,&\qquad
+M_p^2 V_{\phi }\,, \label{eomphi}
\ea
where ``$_{,\phi}=d/d\phi$" and ``$_{,\phi\phi}=d^2/d\phi^2$".


Initially, the universe is slowly expanding (in the Genesis phase),
$H\simeq 0$. We set $V=0$, $g_1= -f_1 e^{2\phi}$, $g_2= f_2$ and
$g_3= f_3$, see e.g. Ref. \cite{Pirtskhalava:2014esa}, with
$f_{1,2,3}$ being dimensionless constants. Thus with Eq. (\ref{eqH}), we
have $\frac{M_p^2}{2} g_1\dot{\phi }^2
  +\frac{3}{4} g_3 \dot{\phi }^4=0$, which suggests
\be
\label{e2phi}
e^{2\phi}={3f_3\over 2M_p^2 f_1}\dot{\phi}^2\,.
\ee
The solution is 
\be \label{dotphi}  \dot{\phi}={1\over (-t)}\,,\quad t<0\,.
\ee

Eq. (\ref{dotH}) reads $\dot{H}={f_3-2f_2\over 4
M_p^2}\dot{\phi}^4$. Thus we get \be \label{HH} H={f_3-2f_2\over
12M_p^2}{1\over (-t)^3}\, \ee after the integration. In principle,
there could be a constant, i.e., $H={f_3-2f_2\over 12M_p^2}{1\over
(-t)^3}+const$, however,  in that case we will have $H\approx
const$ initially, which is geodesically incomplete, see also
\cite{Liu:2013xt}.

Additionally, from Eq. (\ref{HH}), we have  \be a(t)=e^{\int H
dt}=\exp\lf(\frac{ {f_3}-2  {f_2}}{24 M_p^2 t^2}\rt)\simeq
1+\lf(\frac{ {f_3}-2  {f_2}}{24 M_p^2 t^2}\rt)\,, \label{Gat}\ee
while we set $a(-\infty)=1$.

%
%

During inflation, we set $g_1=1$ and $g_2=g_3=0$, since we require
that the inflationary phase is controlled by a simple slow-roll
field\footnote{The behaviors of these $g_i$'s in the two phases
can easily be matched together by making use of some shape
functions \cite{Qiu:2015nha}\cite{Zhang:2006ck}.}.

\subsection{The primordial perturbation and its spectrum}



In the unitary gauge, the quadratic action of the scalar perturbation is presented in the form
of Eq. (\ref{eft_action02}).
The coefficients $c_i$ are (substituting Eqs. (\ref{fff}) to
(\ref{tildem4}) into (\ref{c1})(\ref{c2})({\ref{c3})) \ba
&\,&\label{c11} c_1= {\dot{\phi }^2 \over 4 M_p^2 \gamma^2}
\left[2 \dot{\phi }^2 M_p^2 \left(g_{2,\phi }+2
   g_3\right)-2 g_2 M_p^2 \left(3 H \dot{\phi }+\ddot{\phi} \right)+3 g_2^2
   \dot{\phi }^4\right]
   -{\dot{H} M_p^2\over \gamma^2} \,,
  \\&\,&
c_2=M_p^2\,,
  \\&\,&
c_3={aM_p^2\over \gamma}  Q_{\tilde{m}_4}\,,
\ea
where
\be
\gamma=H+{g_2\over 2M_p^2}\dot{\phi}^3\,,\quad Q_{\tilde{m}_4}=1+{2\tilde{m}_4^2\over M_p^2}\,.
\ee

The sound speed squared $c_s^2$ of scalar perturbation is defined
in Eq. (\ref{C-introduction}). Here, when $\tilde{m}_4^2\equiv 0$ or
$Q_{\tilde{m}_4}=1$,  the sound speed squared of the scalar perturbation is reduced to \ba \label{cs2} c^2_{s0}= 1+ \frac{ 4
\dot{\phi }^2 \left[
 g_2 M_p^2 \lf(\ddot{\phi}- H\dot{\phi } \rt)
+g_2^2 \dot{\phi }^4
+\dot{\phi }^2 M_p^2 \left(g_{2,\phi }+g_3\right)
\right]
   }{
   4 \dot{H} M_p^4+\dot{\phi }^2 \left[ 2 g_2 M_p^2 \left(3 H \dot{\phi }+\ddot{\phi} \right)-3 g_2^2
   \dot{\phi }^4-2 \dot{\phi }^2 M_p^2 \left(g_{2,\phi }+2
   g_3\right)\right]
     }
   \,,
\ea 
It is easy to see that $c_{s0}^2=1$ for inflation, since $g_2=g_3=0$, but not for
Genesis. However, using the operator ${\tilde{m}_4^2(t)\over 2}R^{(3)}\delta
g^{00}$, we could always set $c_s^2=1$ in the Genesis phase,
which requires $\tilde{m}_4^2=-{2M_p^2(f_2+f_3)\over 4f_2+f_3}$.
This suggests $Q_{\tilde{m}_4}=-{3f_3\over 4f_2+f_3}$ is a
constant at $|-t|\gg1$, which is consistent with the solution (\ref{solution22}).



The equation of motion of $\zeta$ is \be u''+\lf({c}_{s}^2
k^2-{z''\over z} \rt)u=0\,, \label{MSEQ} \ee with $u=z\zeta$,
$z=\sqrt{2a^2 c_1}$, the prime denotes the derivative with respect
to the conformal time $\tau=\int dt/a$. The initial state is the
Minkowski vacuum, thus $u= {1\over\sqrt{2c_sk}}e^{-ic_sk\tau}$ for
$\zeta$ modes deep inside the horizon. The power spectrum of
$\zeta$ is  \be P_{\cal R}={k^3\over 2\pi^2}\lf|{u\over z} \rt|^2
\,.\ee



In the following, we will analytically estimate the spectrum of the scalar
perturbation. We set $c_s^2=1$ throughout for simplicity, which
could be implemented by using $Q_{\tilde{m}_4}(t)$, as will be
illustrated by the numerical simulation.





In the Genesis phase, substituting Eqs. (\ref{dotphi}), (\ref{HH}) into
(\ref{c1}), we have \be c_1=\frac{108 f_3 M_p^4}{\left(4
f_2+f_3\right){}^2}(-t)^2\,.\ee Thus \be
 z=
\frac{6  \sqrt{6f_3}  M_p^2}{4 f_2+f_3} (-t)\cdot
\exp\lf(\frac{f_3-2 f_2}{24 M_p^2 t^2}\rt)\,. \ee Then it is
straightforward to obtain ${z''\over z} \approx \frac{\left(f_3-2
f_2\right){}^2}{72 M_p^4 \tau^6}\approx {0\over \tau^2}$, where $
\tau=\int{1\over a}dt \approx t$. Thus the solution of Eq.
(\ref{MSEQ}) is \be u_1={\sqrt{-\pi\tau}\over 2}\lf[C_{11}\cdot
H^{(1)}_{1/2}(-k\tau) +C_{12}\cdot H^{(2)}_{1/2}(-k\tau) \rt]\,,
\ee where  $C_{11}$ and $C_{12}$ are functions of $k$, $H^{(1)}_{\nu}$ and $H^{(2)}_{\nu}$ are the first and
the second kind Hankel functions of $\nu-$th order, respectively.
The initial condition $u= {1\over\sqrt{2k}}e^{-ik\tau}$ indicates
\be C_{11}=i\,,\quad C_{12}=0\,. \ee


In the inflation phase, $c_1=\epsilon M_p^2$, thus $z=\sqrt{2\epsilon
a^2 M_p^2}$. We set $\epsilon\ll 1$ as a constant during
inflation. Then ${z''/ z}\approx(2+3\epsilon)/\tau^2$. The
solution of Eq. (\ref{MSEQ}) is \be u_2={\sqrt{-\pi\tau}\over
2}\lf[C_{21}\cdot H^{(1)}_{\nu_2}(-k\tau) +C_{22}\cdot
H^{(2)}_{\nu_2}(-k\tau) \rt] \ee with $\nu_2\approx 3/2+\epsilon$.

We require that $u_1(\tau_{m})=u_2(\tau_{m})$ and
$u_1'(\tau_{m})=u_2'(\tau_{m})$, with $\tau_m$ approximately
corresponding to the beginning time of inflation phase, and we obtain
\ba &\,&C_{21}=-{i\over4}e^{-ik\tau_m}\sqrt{{-\pi\over 2k \tau_m}}
\lf[ 2k\tau_m H^{(2)}_{\nu_2-1}(-k\tau_m)
+(2\nu_2-1-2ik\tau_m)H^{(2)}_{\nu_2}(-k\tau_m) \rt]\,,
\\
&\,&C_{22}={i\over4}e^{-ik\tau_m}\sqrt{{-\pi\over 2k \tau_m}}
\lf[ 2k\tau_m H^{(1)}_{\nu_2-1}(-k\tau_m)
+(2\nu_2-1-2ik\tau_m)H^{(1)}_{\nu_2}(-k\tau_m) \rt]\,.
\ea
The power spectrum of $\zeta$ is given by \ba P_{\cal R} =P_{\cal
R}^{inf} \cdot|C_{21}-C_{22}|^2\,,\quad k\ll aH\,, \label{PR}\ea
where $P_{\cal R}^{inf}=\frac{H^2_{inf}}{8 \pi ^2 M_p^2 \epsilon
}\cdot \lf({k\over aH} \rt)^{3-2\nu_2}$ is the power spectrum of scalar perturbation modes that exit horizon during inflation. We see that for the
perturbation modes exiting horizon in the Genesis phase, $-k\tau_m\ll
1$, $|C_{21}-C_{22}|^2 \simeq (-k\tau_m)^2$, thus $P_{\cal R}\sim
k^{2}$ is strong blue-tilted, while for the perturbation modes
exiting horizon in the inflation phase, $-k\tau_m\gg 1$,
$|C_{21}-C_{22}|^2 \simeq 1$, thus $P_{\cal R}\sim
k^{3-2\nu_2}=k^{-2\epsilon}$ is flat with a slightly red tilt.


Tensor perturbation is unaffected by the $R^{(3)}\delta g^{00}$
operator. Its quadratic action is given in Eq.
(\ref{tensor-action}) with $Q_T=1$ and $c_T^2=1$. The spectrum of
primordial GWs can be calculated similarly, see also
Ref. \cite{Liu:2014tda}. Since $z_T''/z_T=a''/a$,
we have \be P_T=P_{T}^{inf} \cdot|C_{21}-C_{22}|^2\,,\quad k\ll
aH\,, \ee where $P_{T}^{inf}=\frac{2H^2_{inf}}{\pi ^2 M_p^2
}\cdot \lf({k\over aH} \rt)^{3-2\nu_2}$ is the power spectrum of tensor perturbation modes that exit horizon during inflation. Thus the spectrum of
primordial GWs has a shape similar to that of the scalar perturbation.


\subsection{Numerical simulation}

In the numerical calculation, we set \ba &\,& g_1(\phi)={f_1 e^{2\phi} \over 1+f_1
e^{2\phi} }\tanh\Big[q_1 (\phi-\phi_0) \Big]\,,
\\&\,&
g_{2,3}(\phi)={f_{2,3}}\lf({1-\tanh\Big[q_{2,3}(\phi-\phi_0)\Big]\over
2}\rt)\,,
\\
&\,& V(\phi)=\frac{\lambda}{2} (\phi-\phi_1)^2\lf({1+\tanh
\Big[q_4\left(\phi -\phi_2\right)\Big]\over 2}\rt)  \, \ea with
$f_{1,2,3}$, $q_{1,2,3,4}$, $\phi_{0,1,2}$ and $\lambda$ being
dimensionless constants.  When $\phi\ll \phi_0$, we have $g_1=
-f_1 e^{2\phi}$, $g_2= f_2$ and $g_3= f_3$, which brings a Genesis
phase (\ref{Gat}), while $\phi\gg \phi_0$, we have $g_1=1$ and
$g_2=g_3=0$, the slow-roll inflation will occur with $V(\phi)\sim
\phi^2$. When $\phi\ll \phi_2$, $V(\phi)\approx0$, while $\phi\gg
\phi_2$, $V(\phi)\approx {\lambda\over2}(\phi-\phi_1)^2$. We do
not require $\phi_0=\phi_2$ but $\phi_0>\phi_2$.

We start the simulation at $t_i\ll -1$, and we set \be
\dot{\phi}(t_i)={1\over (-t_i)}\,,\quad
\phi(t_i)={1\over2}\ln\lf[{3f_3\over 2f_1 M_p^2 }{1\over
(-t_i)^2}\rt]\,, \ee and \be a(t_i)=1\,, \quad H(t_i)={f_3-2f_2\over
12M_p^2}{1\over(-t_i)^3}\,. \ee

We show the evolution of $\phi$ and $\dot{\phi}$ in
Fig. \ref{fig001}, and the evolution of $a$, $H$ and $\epsilon$ in
Fig. \ref{fig002}. In Fig. \ref{fig003}(a), $c_1$ is plotted, and
$c_1>0$ is satisfied. In Fig. \ref{fig003}(b), we see that
$\gamma$ does not cross 0, which implies that, in the Genesis phase,
$c_{s0}^2<0$ (see Fig. \ref{fig004}(a)), as proved in Sec.
\ref{dilemma}. By including the operator $R^{(3)}\delta g^{00}$,
we could have $c_s^2>0$, and so cure the gradient instability. The
spectrum of the scalar perturbation can be simulated numerically,
which is plotted in Fig. \ref{fig005}. The spectrum obtained has a
cutoff at large scale $k<k_*$ and is nearly scale-invariant for
$k>k_*$, as displayed in Eq. (\ref{PR}).


\begin{figure}[htbp]
\includegraphics[scale=2,width=0.50\textwidth]{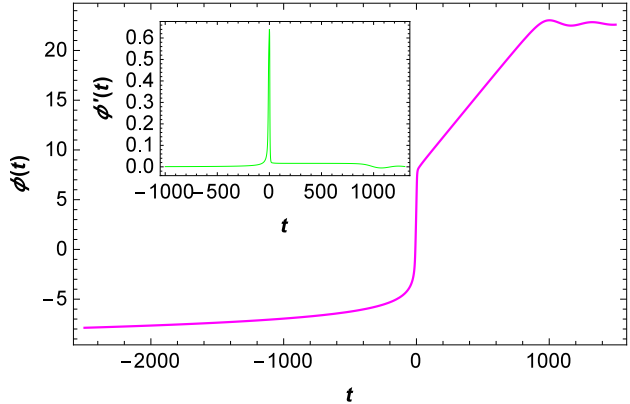}
\caption{~~The evolution of $\phi$ and $\dot \phi$, while we set
$f_1=5$, $f_2=-0.23$, $f_3=-13f_2$, $q_1=1$, $q_2=0.2$, $q_3=0.2$,
$q_4=2$, $\lambda=4\times10^{-4}$, $\phi_0=7$, $\phi_1=22.7$ and
$\phi_2=5.2$. } \label{fig001}
\end{figure}

\begin{figure}[htbp]
\subfigure[~~$a$ and
$H$]{\includegraphics[width=.47\textwidth]{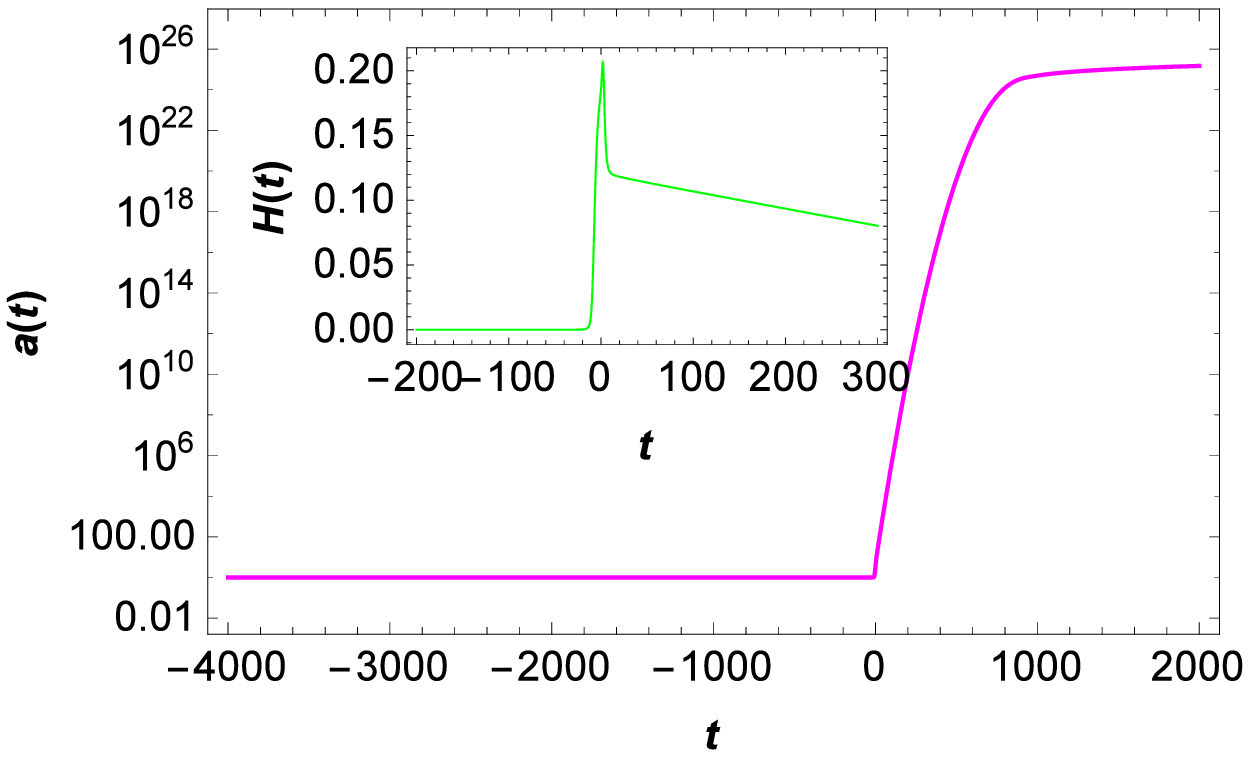} }
\subfigure[~~$\epsilon$]{\includegraphics[width=.44\textwidth]{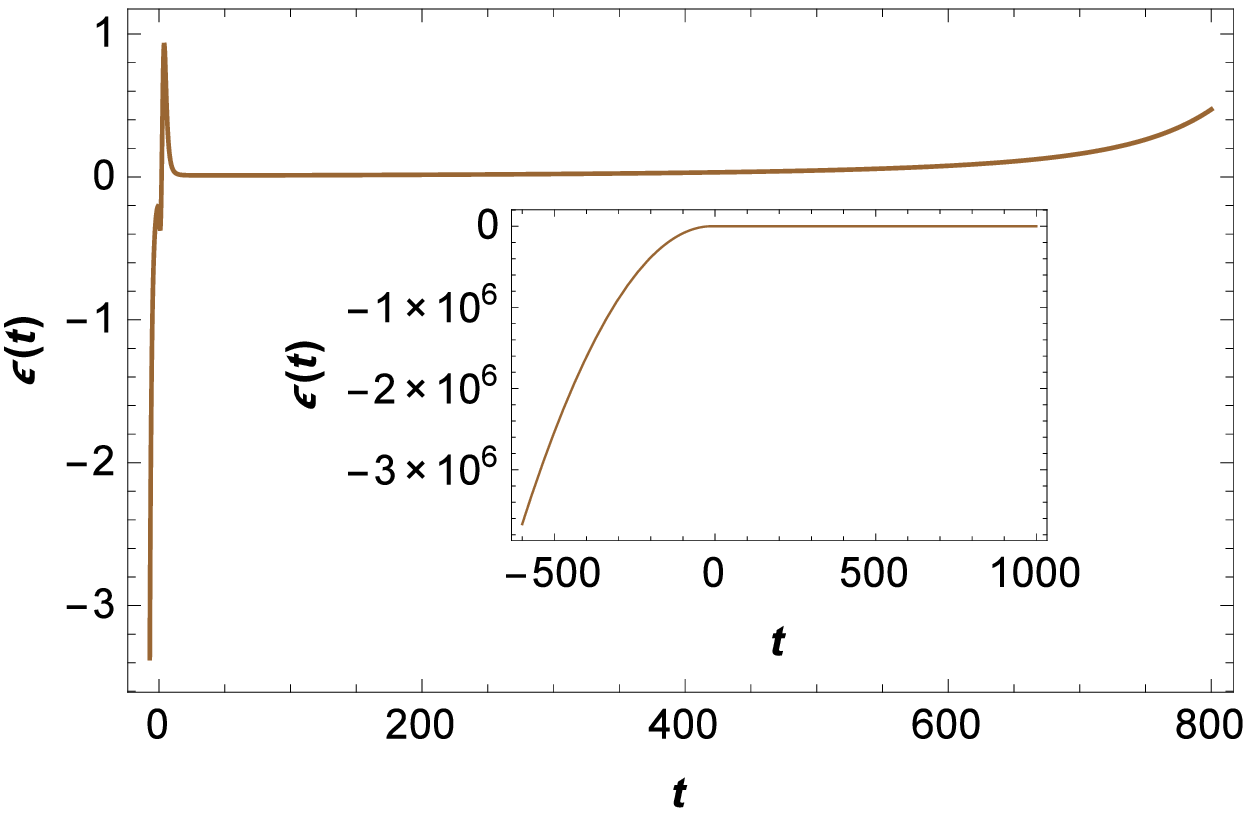}
} \caption{~~The evolution of $a$, $H$ and $\epsilon$, while we
set $f_1=5$, $f_2=-0.23$, $f_3=-13f_2$, $q_1=1$, $q_2=0.2$,
$q_3=0.2$, $q_4=2$, $\lambda=4\times10^{-4}$, $\phi_0=7$,
$\phi_1=22.7$ and $\phi_2=5.2$. } \label{fig002}
\end{figure}



\begin{figure}[htbp]
\subfigure[~~$c_1$]{\includegraphics[width=.47\textwidth]{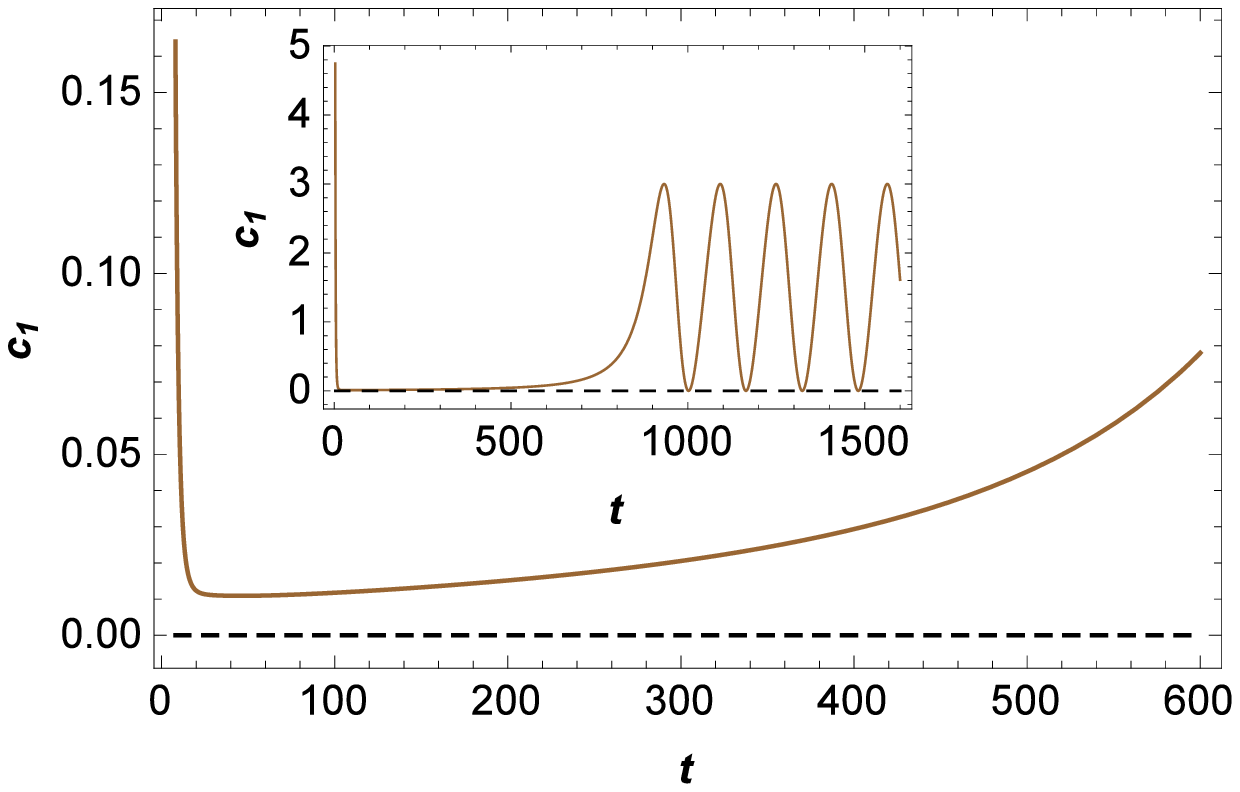}
}
\subfigure[~~$\gamma/H$]{\includegraphics[width=.48\textwidth]{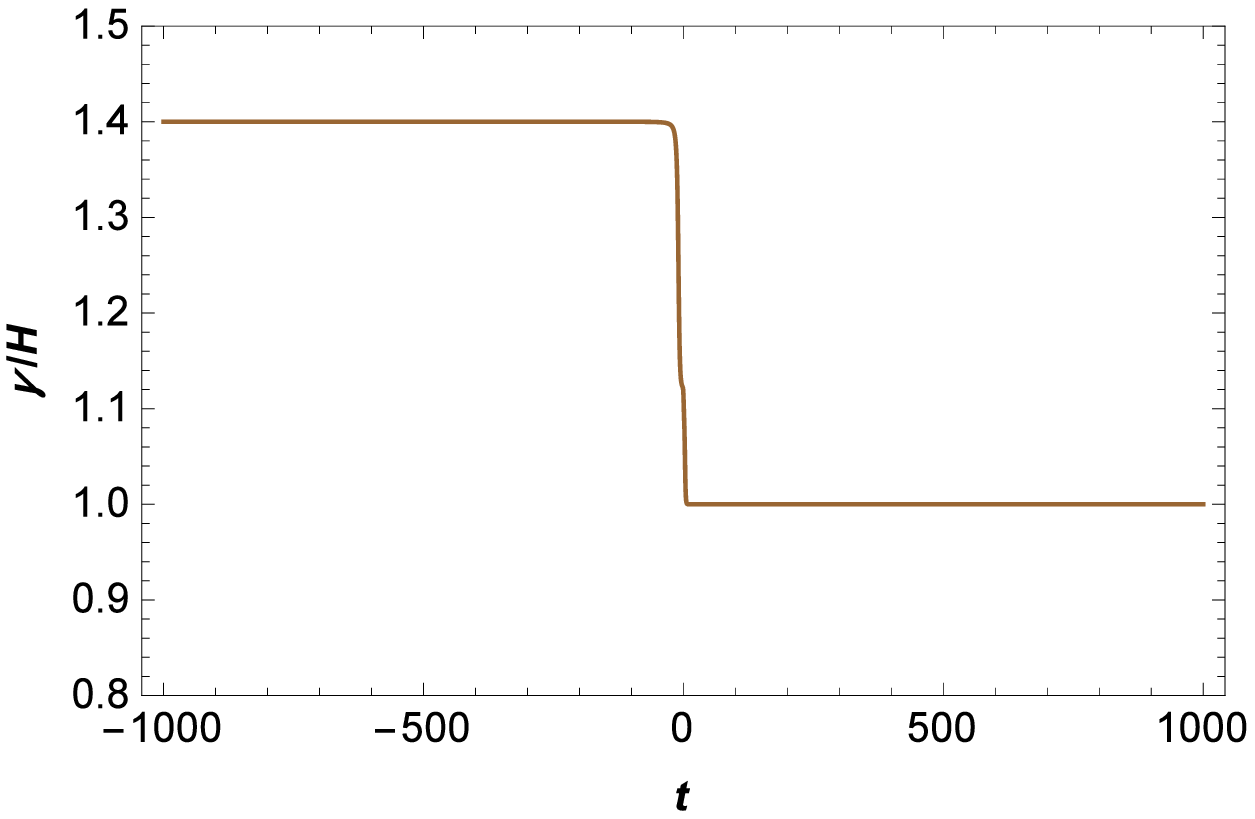}
} \caption{~~The evolution of $c_1$, $\gamma/H$ and $\epsilon$,
while we set $f_1=5$, $f_2=-0.23$, $f_3=-13f_2$, $q_1=1$,
$q_2=0.2$, $q_3=0.2$, $q_4=2$, $\lambda=4\times10^{-4}$,
$\phi_0=7$, $\phi_1=22.7$ and $\phi_2=5.2$. } \label{fig003}
\end{figure}

\begin{figure}[htbp]
\subfigure[~~$c_{s0}^2$ (with $\tilde{m}_4^2\equiv0$) and $c_s^2$
(with
$\tilde{m}_4^2(t)$)]{\includegraphics[width=.47\textwidth]{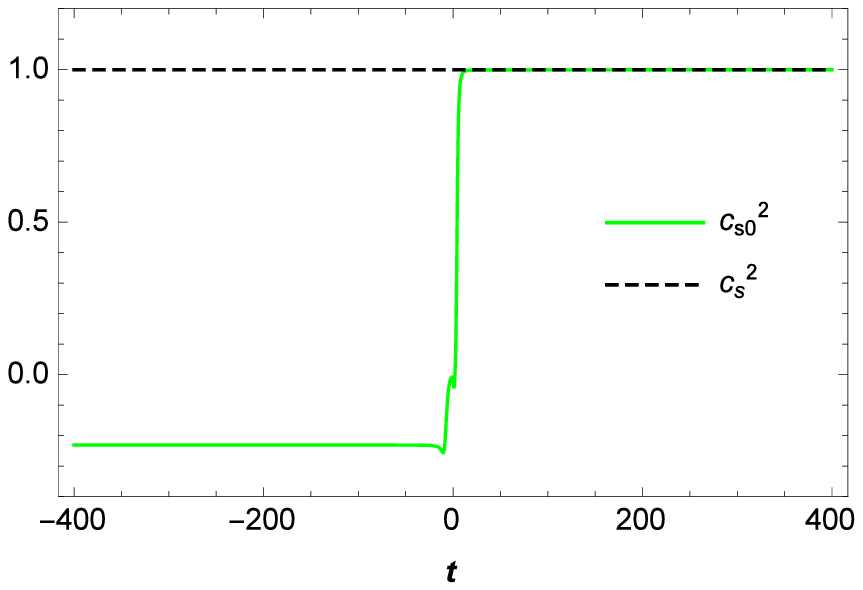}
} \subfigure[~~$Q_{\tilde{m}_4}$ and
$\tilde{m}_4^2$]{\includegraphics[width=.47\textwidth]{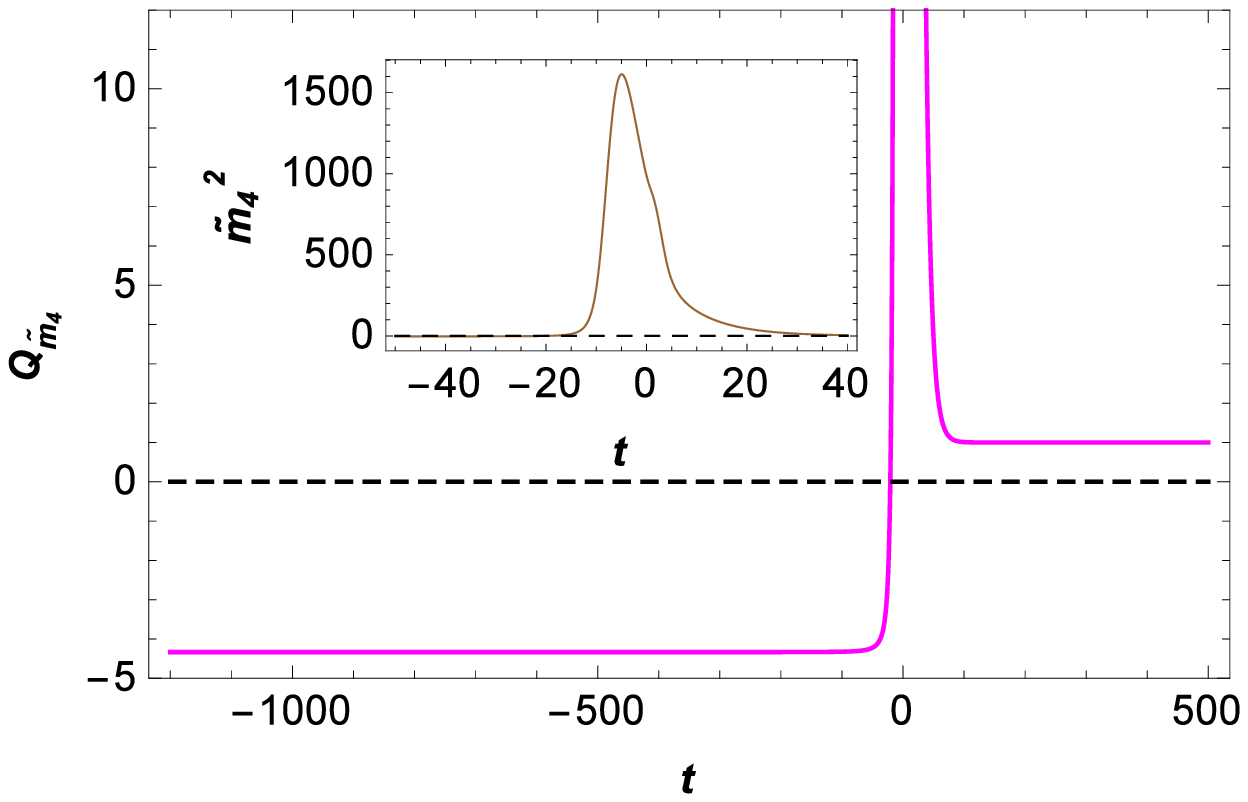}
}
\subfigure[~~$\dot{Q}_{\tilde{m}_4}$]{\includegraphics[width=.47\textwidth]{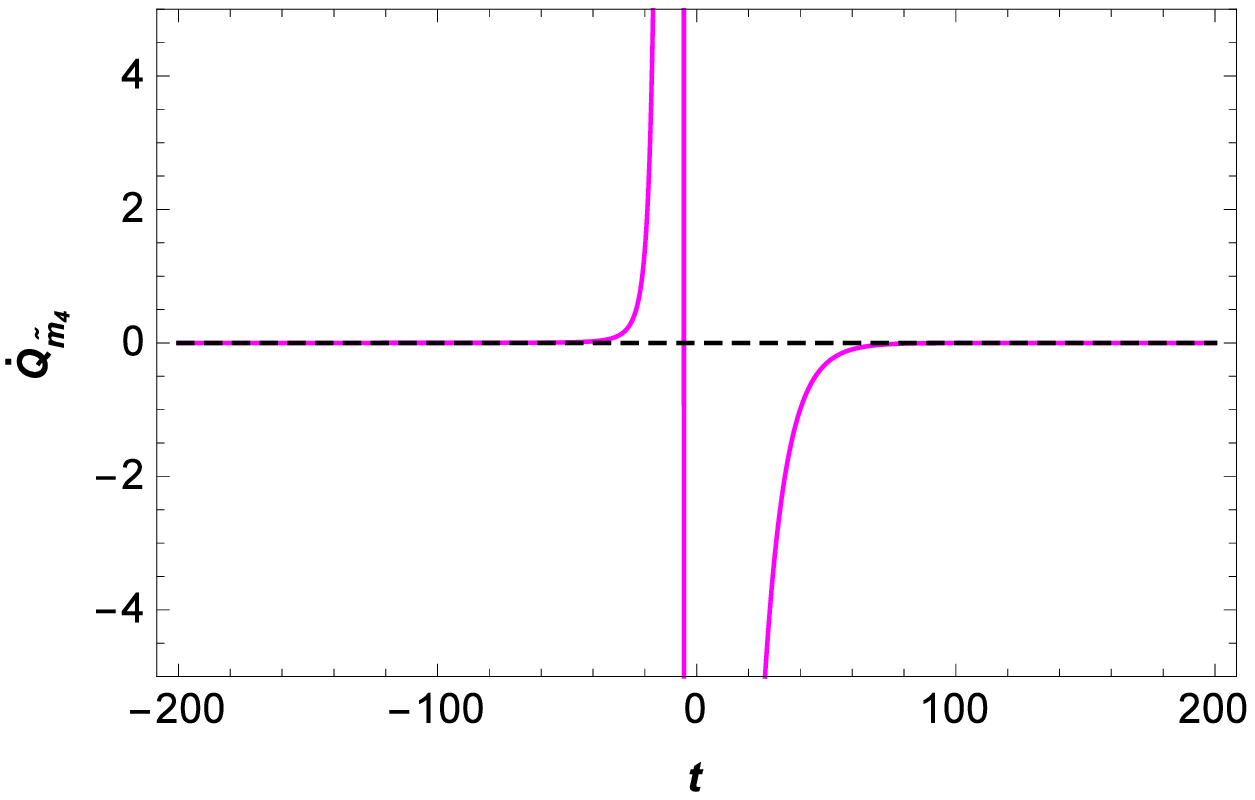}
} \caption{The sound speed of the scalar perturbation with and without
$\tilde{m}_4^2$,  while we set $f_1=5$, $f_2=-0.23$, $f_3=-13f_2$,
$q_1=1$, $q_2=0.2$, $q_3=0.2$, $q_4=2$, $\lambda=4\times10^{-4}$,
$\phi_0=7$, $\phi_1=22.7$ and $\phi_2=5.2$.} \label{fig004}
\end{figure}


\begin{figure}[htbp]
{\includegraphics[width=.47\textwidth]{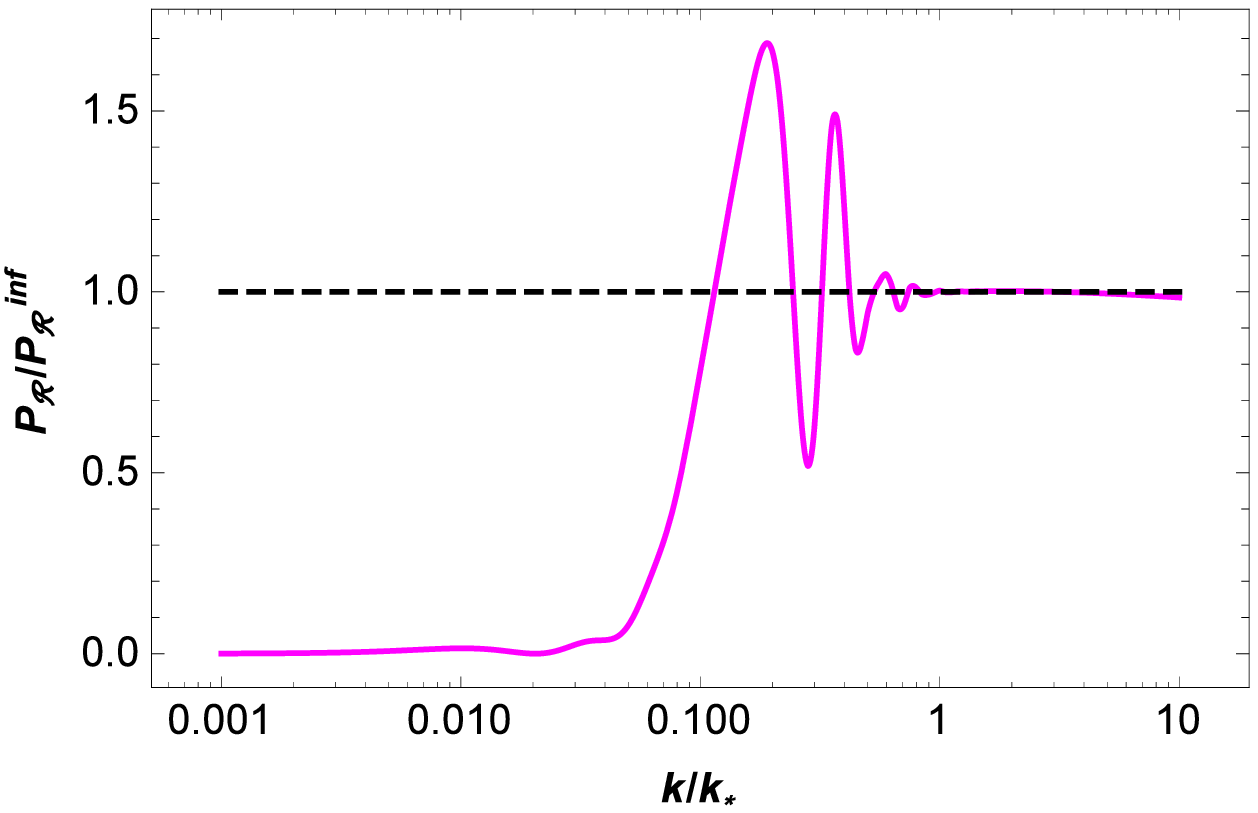} }
\caption{~~The spectrum $P_{\cal R}$ of the scalar perturbation, while
we set $f_1=5$, $f_2=-0.23$, $f_3=-13f_2$, $q_1=1$, $q_2=0.2$,
$q_3=0.2$, $q_4=2$, $\lambda=4\times10^{-4}$, $\phi_0=7$,
$\phi_1=22.7$, $\phi_2=5.2$ and $k_*$ corresponds to the comoving
wave number of the perturbation mode which exits horizon around
the beginning of inflation. } \label{fig005}
\end{figure}



\section{The dilemma of $\gamma$ in the Genesis scenario }\label{dilemma}

In the Genesis scenario based on the cubic Galileon
\cite{Creminelli:2010ba}, (also \cite{Liu:2011ns}), we have \be
\gamma=H+{f_2\over 2M_p^2}\dot{\phi}^3={f_3+4f_2\over
12M_p^2}\dot{\phi}^3\, \label{gammadile}\ee during the Genesis,
where $f_2<0$.

In Ref. \cite{Creminelli:2010ba}, $f_3=-f_2$, which suggests
$\gamma={f_2\over 4M_p^2}\dot{\phi}^3<0$. Thus if a hot ``big
bang" or inflation ($\gamma=H>0$) starts after the Genesis phase,
$\gamma$ must cross 0 at $t_\gamma$ ($c_{s0}^2<0$ around
$t_\gamma$, which may be cured by applying $Q_{\tilde {m}_4}$). It
is obvious that when $\gamma=0$, $c_1$ in (\ref{eft_action02})
will be divergent. Though this divergence might not be a problem,
it will affect the numerical simulation for perturbations
\cite{Battarra:2014tga}\cite{Koehn:2015vvy}, unless
$Q_T/\gamma^2$ is finite at $t_\gamma$, as in Ijjas and
Steinhardt's model \cite{Ijjas:2016vtq}.

In the model of \cite{Pirtskhalava:2014esa}, the Genesis is followed by
Galileon inflation \cite{Kobayashi:2010cm}. Though $f_3=-f_2$ and
$\gamma={f_2\over 4M_p^2}\dot{\phi}^3<0$ in the Genesis phase, one
might also have $\gamma <0$ for Galileon inflation, since
$g_2\neq0$ in (\ref{genesis-action}) during inflation. Thus it
seems that $\gamma$ might not necessarily cross 0. However, after
inflation, $\gamma$ crossing 0 is still inevitable.

In our model, the Genesis is followed by the slow-roll inflation,
$\gamma=H>0$ for inflation. To not cross 0, initially $\gamma$
must satisfy $\gamma>0$. In the Genesis phase, this suggests
$f_3>-4f_2$. Thus we will have $\gamma>0$ throughout. However, for
the cubic Galileon, the expense is \be
c_{s0}^2=1-{4f_2+4f_3\over3f_3}<0\, \ee during the Genesis. Here,
this pathology is cured in EFT by applying (\ref{solution22}).

\section{Conclusion}

Based on the EFT of cosmological perturbations, we revisit the
nonsingular cosmologies, using the ``non-integral approach". By
doing this, we could have a clearer understanding of the pathology
in nonsingular Galileon models and its cure in EFT.


We clarify the application of the operator
${\tilde{m}_4^2}R^{(3)}\delta g^{00}/2$ in EFT, which is significant for curing the gradient instability. We show that if
$Q_{\tilde{m}_4}<0$ around $\gamma=0$ is adopted to cure the gradient instability,  in solution (\ref{solution12}) (with
$\gamma<0$ and ${Q}_{\tilde{m}_4}=1$ initially),
${Q}_{\tilde{m}_4}$ must cross 0 twice; while in solution
(\ref{solution22}) (with $\gamma>0$ throughout), initially
$Q_{\tilde{m}_4}<0$ must be satisfied, $Q_{\tilde{m}_4}$ will
cross 0 to $Q_{\tilde{m}_4}>0$ at $t_{\tilde{m}_4}$, and crosses 0
only once. Thus at a certain time, ${Q}_{\tilde{m}_4}$ meeting 0 is
required, as pointed out first by Cai et.al \cite{Cai:2016thi},
and also by Creminelli et.al \cite{Creminelli:2016zwa}.

We also clarify that in the bounce model with $\gamma<0$ initially,
$c_s^2<0$ will occur in the phase with $\gamma\simeq 0$ and ${\dot
\gamma}>0$, while the NEC is violated when ${\dot H}>0$ (bounce
phase), these two phases do not necessarily coincide. As pointed out by
Ijjas and Steinhardt \cite{Ijjas:2016vtq}, it is the sign's change of
$\gamma$ that causes $c_s^2<0$. Here, we verify this point.
In Genesis model
\cite{Creminelli:2010ba}\cite{Pirtskhalava:2014esa}, and also
\cite{Liu:2011ns}, the case is similar, as discussed in Sec.
\ref{dilemma}.

The nonsingular model with the solution (\ref{solution22})
($\gamma>0$ throughout) has not been  studied before. In Sec.
\ref{gene-inf}, we design such a model, in which a slow expansion phase (namely, the Genesis phase) 
is followed by slow-roll inflation. Under the unitary gauge, since
${\dot \gamma}>0$ and $\gamma>0$ (not crossing 0), the evolution
of primordial perturbations can be simulated numerically. The
simulation displays that the spectrum acquires a large-scale cutoff, as
expected in Ref. \cite{Liu:2014tda}.

We conclude that, based on EFT, not only a 
stable
nonsingular cosmological scenario may be built without getting involved in
unknown physics, but also the phenomenological possibilities of
its implementation are far richer than expected (see also recent
\cite{Giovannini:2016jkf}\cite{Misonoh:2016btv} for the higher
spatial derivative operators).



\textbf{Acknowledgments}

We thank Youping Wan for helpful comments. The work of YSP is supported
by NSFC, Nos. 11575188, 11690021, and also supported by the
Strategic Priority Research Program of CAS, Nos. XDA04075000,
XDB23010100.
The work of T. Q. is supported by NSFC under Grant Nos. 11405069 and 11653002.

\appendix

\section{EFT of cosmological perturbations}
\label{EFT}

With the ADM line element, we have
\begin{equation}
g_{\mu\nu}=\left(
  \begin{array}{cc}
  N_kN^k-N^2 &  N_j\\
  N_i &  h_{ij}\\
  \end{array}
\right) \,,\qquad
g^{\mu\nu}=\left(
  \begin{array}{cc}
  -N^{-2} &  {N^j\over N^2}\\
  {N^i\over N^2} &  h^{ij}-{N^iN^j\over N^2}\\
  \end{array}
\right) \,,\qquad
\end{equation}
and $\sqrt{-g}=N\sqrt{h}$, where $N_i=h_{ij}N^j$. The unit
one-form tangent vector is defined as $n_{\nu}=n_0
(dt/dx^{\mu})=(-N,0,0,0)$ and $n^{\nu}=g^{\mu\nu}n_{\mu} =({1/
N},-{N^i/ N})$, which satisfies $n_{\mu}n^{\mu}=-1$. On the
hypersurface, the induced 3-dimensional metric is
$H_{\mu\nu}=g_{\mu\nu}+n_{\mu}n_{\nu}$, thus
\begin{equation}
H_{\mu\nu}=\left(
  \begin{array}{cc}
  N_kN^k &  N_j\\
  N_i &  h_{ij}\\
  \end{array}
\right) \,,\qquad
H^{\mu\nu}=\left(
  \begin{array}{cc}
  0 &  0\\
  0 &  h^{ij}\\
  \end{array}
\right) \,.\qquad
\end{equation}
The extrinsic curvature is $K_{\mu\nu}\equiv{1\over2}{\cal
L}_{n}H_{\mu\nu}$, where ${\cal L}_{n}$ is the Lie derivative with
respective to $n^{\mu}$. The induced 3-dimensional Ricci scalar
$R^{(3)}$ associated with $H_{\mu\nu}$ is
\be \label{Ricci}
R^{(3)}=R +K^2-K_{\mu\nu}K^{\mu\nu}-2\nabla_\mu(Kn^\mu-n^\nu\nabla_\nu
n^\mu)~.
\ee

Without higher-order spatial derivatives, the EFT reads
\cite{Cai:2016thi} \ba\label{eft_action} S&=&\int
d^4x\sqrt{-g}\Big[ {M_p^2\over2} f(t)R-\Lambda(t)-c(t)g^{00}
\nn\\
&\,&+{M_2^4(t)\over2}(\delta g^{00})^2-{m_3^3(t)\over2}\delta
K\delta g^{00} -m_4^2(t)\lf( \delta K^2-\delta K_{\mu\nu}\delta
K^{\mu\nu} \rt) \nn\\ &\,& + {\tilde{m}_4^2(t)\over
2}R^{(3)}\delta g^{00}\Big] +S_m[g_{\mu\nu},\psi_m]\,,
\label{action}\ea 
where $\delta g^{00}=g^{00}+1$, $\delta K_{\mu\nu}=K_{\mu\nu}-H_{\mu\nu}H$ with $H$ being the Hubble parameter.
The coefficient set $(f, c, \Lambda, M_2, m_3$,
$m_4, {\tilde m}_4)$ specifies different theories and could be time-dependent in general \footnote{Different conventions of the nomenclatures of these coefficients were adopted during the development of the EFT of cosmological perturbations (see e.g., \cite{Cheung:2007st}\cite{Gubitosi:2012hu}\cite{Piazza:2013coa}\cite{Gleyzes:2013ooa}). Here, we follow the convention used in Refs. \cite{Piazza:2013coa}\cite{Gleyzes:2013ooa}.}. A particular
subset $(m_4= {\tilde m}_4)$ of EFT (\ref{action}) is the
Horndeski theory. $S_m[g_{\mu\nu},\psi_m]$ is the matter part,
which is minimally coupled to the metric $g_{\mu\nu}$.

To obtain the quadratic actions for scalar and tensor perturbations, we will work in the unitary gauge, thus we set
\be h_{ij}=a^2e^{2\zeta}(e^{\gamma})_{ij},\qquad \gamma_{ii}=0=\partial_i\gamma_{ij} \,.
\ee
Then we follow the standard method first used by Maldacena \cite{Maldacena:2002vr}, it is straightforward (though tedious) to obtain the quadratic actions of  scalar perturbation $\zeta$ and tensor perturbation $\gamma_{ij}$, as exhibited in Eqs. (\ref{eft_action02}) and (\ref{tensor-action}), respectively (see \cite{Cai:2016thi} for detailed derivations).

 \end{document}